\documentclass[letter]{jpsj2} %% for letters
\title{A Kondo lattice antiferromagnet CePd$_{5}$Al$_{2}$}

\author{R. A. Ribeiro$^{1}$\thanks{E-mail address: ribeiro@hiroshima-u.ac.jp},
T. Onimaru$^{2}$,
K. Umeo$^{3}$,
M. A. Avila$^{2}$,
K. Shigetoh$^{2}$ and
T. Takabatake$^{1,2}$\thanks{E-mail address: takaba@hiroshima-u.ac.jp}}

\inst{$^{1}$ Institute for Advanced Materials Research, Hiroshima University, Higashi-Hiroshima 739-8530\\
$^{2}$ Department of Quantum Matter, ADSM, Hiroshima University, Higashi-Hiroshima 739-8530\\
$^{3}$ Cryogenics and Instrumental Analysis Division, N-BARD, Hiroshima University, Higashi-Hiroshima 739-8526\\}

\abst{We report on the electrical resistivity, magnetic
susceptibility and heat-capacity measurements on a new
intermetallic compound CePd$_{5}$Al$_{2}$, crystallizing in the
ZrNi$_{2}$Al$_{5}$-type tetragonal structure, with lattice
parameters $a = 4.156$~\AA ~and $c = 14.883$~\AA. The compound
presents Kondo lattice behavior and an easy-plane
antiferromagnetic ground state with two magnetic transitions at
2.9~K and 3.9~K. The Sommerfeld coefficient is estimated as
60~mJ/mol~K$^{2}$.}

\kword{cerium palladium aluminum, heavy fermion, Kondo lattice,
antiferromagnetism}

\begin{document}
\maketitle

%% No sections necessary for express letters, letters and short notes

Cerium-based intermetallic compounds have been intensively studied
due to the exceptional variety of physical properties they can
display, and their potential to be used as model systems for
understanding fundamental phenomena such as quantum critical
points or unconventional superconductivity \cite{settai}. For
example, among several compounds in the ternary system Ce-Ni-Al,
CeNi$_{2}$Al$_{5}$ and CeNiAl$_{4}$ have attracted attention due
to their interesting electric and magnetic properties
\cite{Isikawa01,Mizushima01,Mizushima02,Isikawa02,Isikawa03,Isikawa04,Kashiwakura}.
In fact, CeNi$_{2}$Al$_{5}$ is a heavy fermion compound with a
magnetic transition at 2.6~K and a Kondo temperature $T_{K} \simeq
5$~K. On the other hand, CeNiAl$_{4}$ is a non-magnetic heavy
fermion compound with $T_{K} \simeq 67$~K. In the Ce-Pd-Al system,
the hexagonal compounds CePdAl and CePd$_{2}$Al$_{3}$ order
antiferromagnetically below  2.7~K and 2.8~K, respectively. In the
latter, a large Sommerfeld coefficient $\gamma = 340$~mJ/
mol~K$^{2}$ was observed \cite{xue,kitazawa94,kitazawa92}.
Recently, a new transuranic compound NpPd$_{5}$Al$_{2}$ has been
reported \cite{aoki07}, a rare example of a Np compound without
magnetic ordering, which also presents a superconducting
transition temperature at $T_{C} = 4.9$~K. Motivated by these
works, it seemed worthwhile to investigate the existence of a new
Ce compound CePd$_{5}$Al$_{2}$, which we have succeeded to prepare
and herein report the physical properties.

\begin{figure}[b]
\begin{center}
\includegraphics[width=160mm]{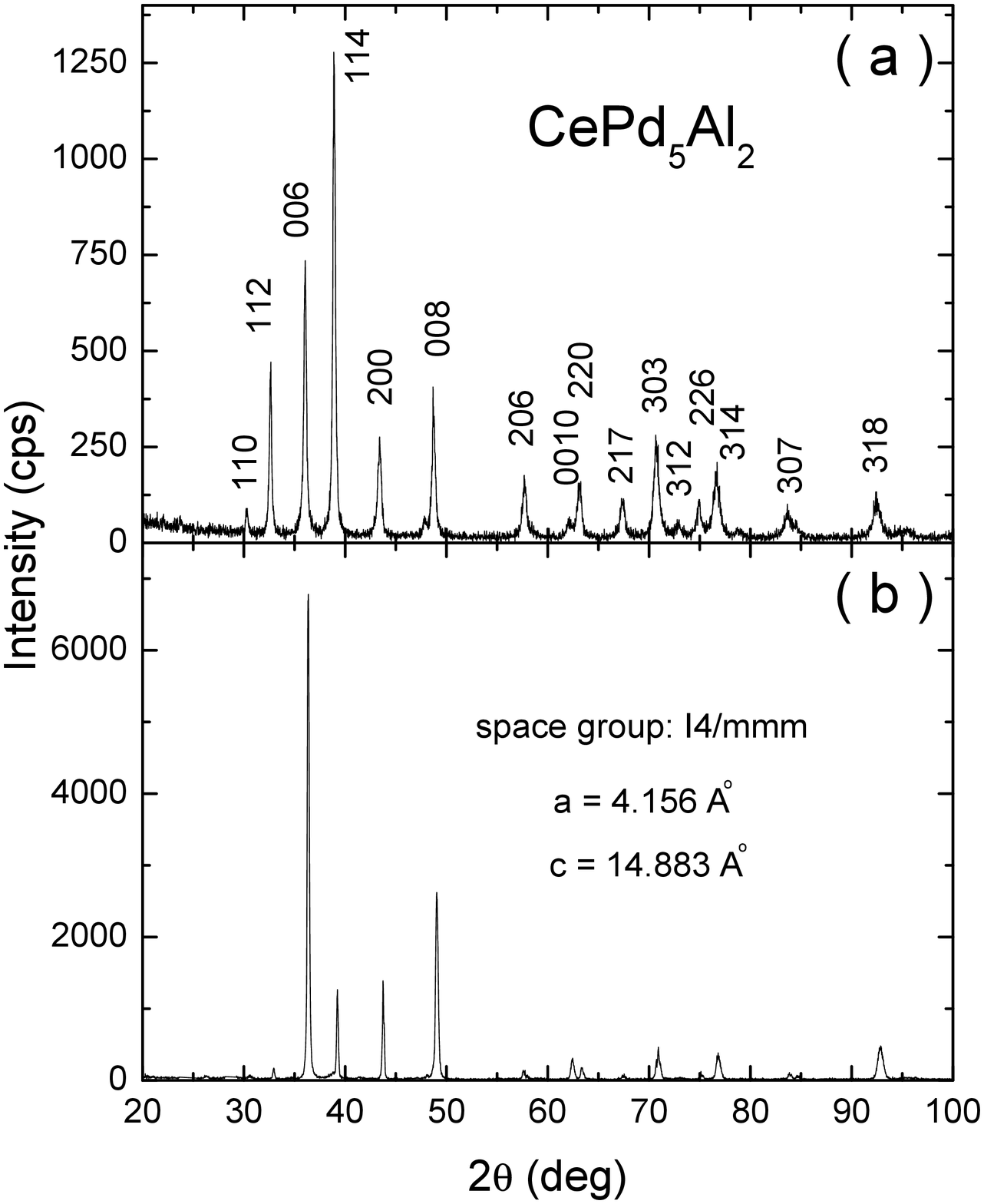}
\caption{X-ray diffraction (Cu $K\alpha$) pattern of
CePd$_{5}$Al$_{2}$ at room temperature. (a) powdered sample, (b) bottom surface of the arc melted button.} \label{xrays}\end{center}
\end{figure}

A polycrystalline sample of CePd$_{5}$Al$_{2}$ was prepared by arc
melting in a high purity argon atmosphere. The initial mixture was
made directly from the elements (Ce: 99.9\%, Pd: 99.99\%, Al:
99.999\%) in stoichiometric proportion and melted into a button.
After that, it was sealed in an a evacuated quartz tube and
annealed at 800$^{\circ}$C for 7 days. The chemical composition of
the annealed button was evaluated by electron-probe microanalysis
(EPMA) using a wavelength dispersive JEOL JXA-8200 system. A
stoichiometric 1:5:2 proportion in the CePd$_{5}$Al$_{2}$ sample
was confirmed and no secondary phases were found. Differential
thermal analysis up to 1550$^{\circ}$C revealed a single
exothermal event at 1080$\pm10^{\circ}$C, associated the
compound's melting point. Powder x-ray diffraction also revealed
that the sample presents a single phase, as shown in
Fig.~\ref{xrays}. The phase is isostructural to
NpPd$_{5}$Al$_{2}$: a tetragonal ZrNi$_{2}$Al$_{5}$-type
structure, with space group $I4/mmm$. The lattice constants
obtained by refining the x-ray pattern in the Powder Cell program
are $a = 4.156$~\AA ~and $c = 14.883$~\AA. We could also observe
in metallographic examination that the grains had an elongated
shape of 1~mm in length and 100~$\mu$m in diameter, indicative of
strongly preferential alignment. An x-ray diffraction pattern
(Fig.~\ref{xrays}.b) from the polished bottom of the sample showed
high intensities for the $00l$ peaks, indicating that the c-axis
is preferentially aligned with the long direction of the grains. A
sample of LaPd$_{5}$Al$_{2}$ was prepared by the same method as
for Ce compound, in attempt to obtain a non-magnetic reference
compound. However, multiple phases were observable in the powder
x-ray diffraction so this reference sample was not used in this
work.

Based on the preferential alignment, elongated bar shaped samples
were cut from the CePd$_{5}$Al$_{2}$ pellet parallel and
perpendicular to the grain length for resistivity experiments. In
addition, a cube shaped sample was cut for magnetization
experiments parallel and perpendicular to this direction, and a
platelike sample was cut for heat capacity experiments. The
electrical resistivity $\rho(T)$ was measured by a AC four-probe
method in two different systems: a $^3$He refrigerator for
temperatures between 0.3~K to 100~K, and a GM refrigerator
home-built setup for 3~K to 400~K. The magnetization $M(T,H)$ was
measured by a SQUID magnetometer (Quantum Design MPMS) from 2 to
350~K. The specific heat $C(T)$ was also measured using a Quantum
Design PPMS, between 0.3~K and 300~K.

\begin{figure}[b]
\begin{center}
\includegraphics[width=160mm]{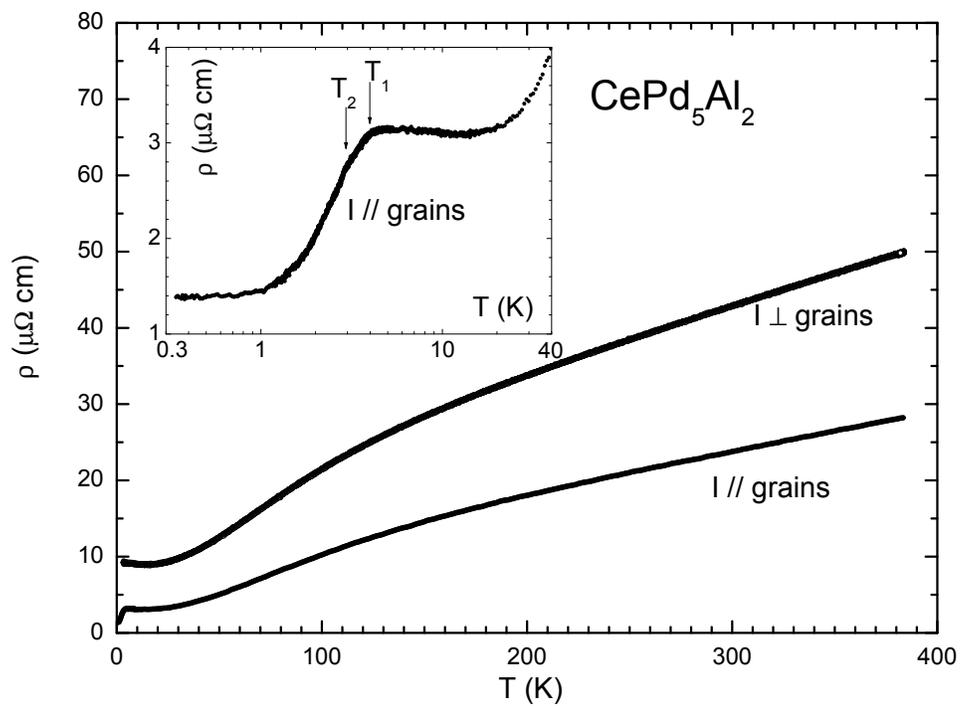}
\caption{Temperature dependence of the electrical resistivity,
$\rho(T)$, of CePd$_{5}$Al$_{2}$. The inset shows the $\rho(T)$ vs $\ln T$
plot and the arrows mark the two anomalies.}
\label{rho}
\end{center}
\end{figure}

Fig.~\ref{rho} shows the temperature dependence of electrical
resistivity, $\rho(T)$, of CePd$_{5}$Al$_{2}$ with the current
parallel and perpendicular to the grain alignment. The sample
shows anisotropy and classic strongly correlated electron metal
behavior \cite{hewson} with
$\rho_{\parallel}$(300~K)~=~24~$\mu\Omega$~cm and
$\rho_{\perp}$(300~K)~=~42~$\mu\Omega$~cm, decreasing as a broad
shoulder upon cooling until a broad local minimum is reached at 13
K. Kondo lattice behavior is then clearly seen (inset of
Fig.~\ref{rho}), where a $-\ln T$ dependence appears between 10~K
and 4~K. A drop in the $\rho(T)$ at $T_{1} = 3.9$~K is ascribable
to the reduction of the scattering of conduction electrons due the
onset of Ce magnetic moments coherence and/or alignment. A second
kink in the slope at $T_{2} = 2.9$~K is also seen, marking a
probable change in magnetic structure. Fermi-liquid behavior can
be observed below 0.8~K and a $\rho = \rho_{0} + AT^{2}$ fit below
this temperature gives residual resistivity $\rho_{0} =
1.37$~$\mu\Omega$~cm and $A = 0.04$~$\mu\Omega$~cm/K$^{2}$. The
residual resistivity ratio ($RRR = \rho(300$~K$) / \rho(0$~K$))$
is 18, which demonstrates a good quality of the sample.

\begin{figure}[b]
\begin{center}
\includegraphics[width=160mm]{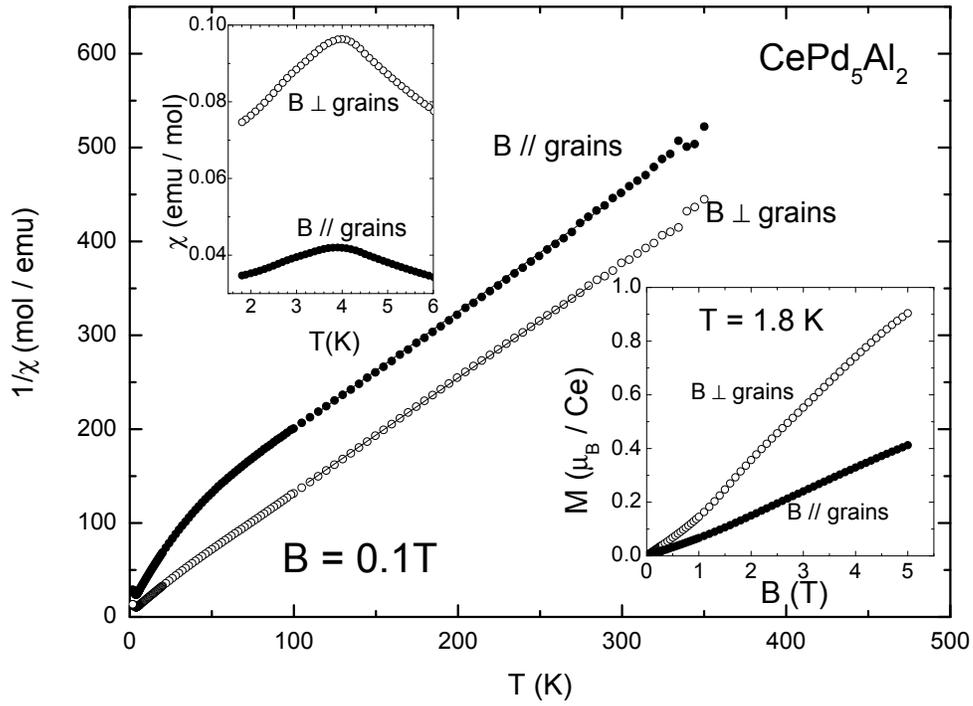}
\caption{Temperature dependence of the inverse magnetic
susceptibility $1/\chi = B/M$ for CePd$_{5}$Al$_{2}$ at $B =
0.1$~T, showing magnetocrystalline anisotropy and Curie-Weiss
behavior at high temperatures. The solid circles and open circles
are the measurements with $B$ parallel and perpendicular to the
grain lengths, respectively. The upper inset shows the low
temperatures data of $\chi (T)$. The lower inset shows the
magnetization isotherms at $T = 1.8$~K, revealing a metamagnetic
behavior at 0.9~T for both orientations.} \label{M}
\end{center}
\end{figure}

The temperature dependence of the reciprocal susceptibility,
$1/\chi = B/M$ at applied field $B = 0.1$~T is shown in
Fig.~\ref{M}. Since the CePd$_{5}$Al$_{2}$ polycrystalline sample
presented grain alignment, we could perform magnetization
measurements in both directions, parallel and perpendicular to the
grain length, corresponding to the solid and open circles in
Fig.~\ref{M}, respectively. The measurements indicate the presence
of easy-plane magnetocrystalline anisotropy, with larger response
perpendicular to the alignment of the grains, therefore to the
$c$-axis. Curie-Weiss law fittings between 175~K and 275~K of the
two datasets are shown as straight lines in the figure. For
$\chi_{\parallel}$, the Curie temperature is $\theta_{\parallel} =
-59$~K, while for $\chi_{\perp}$ we obtained $\theta_{\perp} =
-8.1$~K. The estimated effective moments are very similar,
$\mu_{eff} = 2.53$~$\mu_{B}$/f.u. and $2.56$~$\mu_{B}$/f.u.
respectively, indicating that in CePd$_{5}$Al$_{2}$ the Ce ions
are trivalent. The large difference in $\theta$ suggests a
crystalline electric field (CEF) splitting of the Ce ground state,
and from the CEF model\cite{cho} we can roughly estimate the value
of $B_{2}^{0} = \{[10 (\theta_{\perp} - \theta_{\parallel})] /
[3(2J - 1)(2J + 3)]\}$ as 5.3~K, but a single crystal will be
necessary to do a more precise evaluation. In both directions, an
antiferromagnetic transition with $T_{N1} = 3.9$~K is found,
followed by a second magnetic transition at $T_{N2} = 2.9$~K. The
inset in Fig.~\ref{M} shows the magnetization isotherms at 1.8~K
for the parallel and perpendicular directions of the grains,
revealing a metamagnetic behavior at $B \simeq 0.9$~T.

\begin{figure}[b]
\begin{center}
\includegraphics[width=160mm]{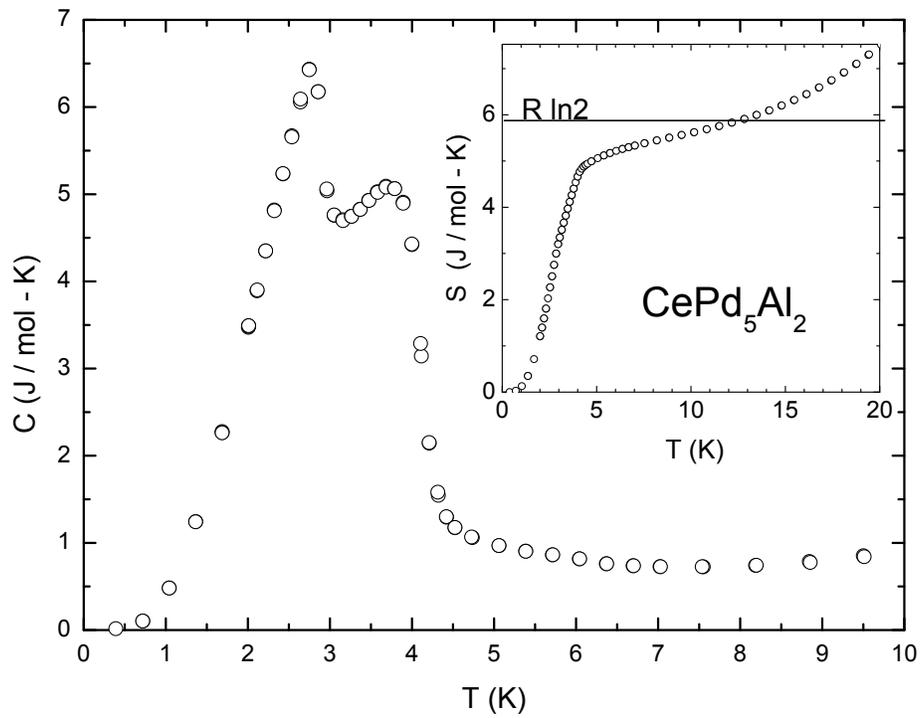}
\caption{Temperature dependence of the specific heat $C(T)$ for
CePd$_{5}$Al$_{2}$, showing a double peak structure at low
temperatures. The inset shows the total entropy $S(T)$.} \label{C}
\end{center}
\end{figure}

The specific heat $C(T)$ measured at zero field is presented in
Fig.~\ref{C}. Two very distinct peaks are seen, giving $T_{N1} =
3.9$~K and $T_{N2} = 2.9$~K, consistent with the double magnetic
transitions found in resistivity and magnetization data. By
extrapolating a linear behavior in $C/T$ versus $T^{2}$ data
between 8~K and 20~K (not shown) the Sommerfeld coefficient of
CePd$_{5}$Al$_{2}$ is estimated as $\gamma = 60$~mJ/mol~K$^{2}$.
This not only provides information about the density of states at
the Fermi level for a metal, it is also a good parameter to
evaluate the strength of electronic correlations. In a classical
metal $\gamma$ is typically rather small ($\sim
10$~mJ/mol~K$^{2}$), while true heavy-fermion systems have values
higher than 400~mJ/mol~K. In CePd$_{5}$Al$_{2}$ the intermediate
$\gamma$ value indicates that electronic correlations are present,
but it cannot be considered a true heavy-fermion compound.
Integration of a $C/T$ versus $T$ plot gives an estimation of the
total entropy $S$ (inset of Fig.~\ref{C}) which includes magnetic,
electronic and lattice contributions. The total entropy of
CePd$_{5}$Al$_{2}$ does not reach $R\ln2$ at the $T_{N1}$, and
does only well above the magnetic transitions, at $T \approx
13$~K. This is further indication of electron correlations and
magnetic screening effects, i.e., the $4f$ electrons of the Ce
ions do not behave as simple localized electrons. The CEF ground
state of the Ce ions should be a doublet, similar to the
CeNi$_{2}$Al$_{5}$ compound \cite{aoki07}. The Kadowaki-Woods
ratio $A/\gamma^2$ results in the expected value of
$10^{-5}~\Omega$~cm~(mol~K/J)$^2$, also supporting a ground-state
doublet \cite{tsujii,kontani}.

In summary, we have prepared a new compound CePd$_{5}$Al$_{2}$,
which presents two magnetic transitions at $T_{N1} = 3.9$~K and
$T_{N2} = 2.9$~K. The resistivity measurements reveal Kondo
lattice and Fermi-liquid behavior. The compound has anisotropic
magnetization and presents a CEF splitting of the Ce ground state,
therefore we are now making efforts to grow single crystals for
further investigations.

\section*{Acknowledgment}

We thank Y. Shibata for the electron-probe microanalysis performed at N-BARD, Hiroshima University. This work was financially supported by the Grant in Aid for Scientific Research (A), No. 18204032, from Ministry of Education, Culture, Sports, Science and Technology.

\clearpage
\vspace{50 mm}

\clearpage
\vspace{50 mm}

\clearpage
\vspace{50 mm}

\clearpage
\vspace{50 mm}

\end{document}